# Imaging carrier transport properties in halide perovskites using time-resolved optical microscopy

*Géraud Delport, Stuart Macpherson and Samuel D. Stranks\**

Dr. Géraud Delport, Stuart Macpherson, Dr. Samuel D. Stranks, Cavendish Laboratory, University of Cambridge, JJ Thomson Avenue, Cambridge CB3 0HE, UK. *E-mail: sds65@cam.ac.uk

Dr. Samuel D. Stranks, Department of Chemical Engineering & Biotechnology, University of Cambridge, Philippa Fawcett Drive, Cambridge CB3 0AS, UK

**Halide perovskites have remarkable properties for relatively crudely processed semiconductors, including large optical absorption coefficients and long charge carrier lifetimes. Thanks to such properties, these materials are now competing with established technologies for use in cost-effective and efficient light harvesting and light emitting devices. Nevertheless, our fundamental understanding of the behaviour of charge carriers in these materials – particularly on the nano-to micro-scale – has on the whole lagged behind the empirical device performances. Such understanding is essential to control charge carriers, exploit new device structures, and push devices to their performance limits. Among other tools, optical microscopy and spectroscopic techniques have revealed rich information about charge carrier recombination and transport on important length scales. In this Progress Report, we detail the contribution of time-resolved optical microscopy techniques to our collective understanding of the photophysics of these materials. We discuss ongoing technical developments in the field that are overcoming traditional experimental limitations in order to visualise transport properties over multiple time and length scales. Finally, we propose strategies to combine optical microscopy with complementary techniques in order to obtain a holistic picture of local carrier photophysics in state-of-the-art perovskite devices.**

Over the last decade, there has been a dramatic rise in the efficiencies of light harvesting[1,2], and light emitting[3,4] devices based on halide perovskite materials, with performances already rivalling existing commercial technologies. Perovskites are high-quality semiconductors with remarkable optoelectronic properties including an apparent defect tolerance[5], long charge-carrier lifetimes[6], long charge-carrier diffusion lengths[7,8] and sharp absorption edges[9]. However, they also exhibit complex heterogeneous morphological, chemical, structural and photo-physical properties across multiple length scales[10] arising from combinations of their hybrid organic-inorganic nature[11,12], mixed chemical compositions[13] and their polycrystalline structure[14], presenting challenges for rigorous charaterisation. A broad range of experimental techniques[10] has been employed to provide insight into their structural, chemical, morphological and macroscopic device operation properties, including electrical devices, diffraction, and electron microscopy characterization approaches, but these tools fall short in elucidating critical processes impacting device operation, namely the dynamics of photo-excited species and the transport of energy on all length scales. Optical spectroscopy techniques [15–19] directly probe the photophysics of materials and devices (see Figure 1a and b) making them powerful tools to establish and contextualise important properties such as absorption and photoluminescence[15] (PL) spectra[20,21], exciton binding energies[22], PL or device quantum efficiencies[4,23], defect concentrations[24,25] and the degree of sub-gap disorder[9,26] (via measurement of the Urbach tail). In particular, techniques employing time-resolved optical spectroscopy coupled with optical microscopes allow us to probe local photophysical phenomena such as the recombination and transport of charge carriers over a range of temporal (from seconds to femtoseconds) and spatial (from centimetres to nanometres) scales (see summary in Figure 1a). In this Progress Report, we will outline the use of these local spectroscopic techniques in revealing critical information about charge carrier behaviour in halide perovskite materials. We will discuss key capabilities and limitations of these approaches to ascertain important device-relevant information. Finally, we will highlight the need to exploit *in operando* optical microscopy techniques concomitant with multimodal microscopy approaches to gain a broad perspective of these complex local phenomena under true device operating conditions. Such developments will allow us to uncover the fundamental science that will directly steer future device improvements, reversing the way the field has primarily proceeded to date whereby empirical device improvements have preceded fundamental studies.

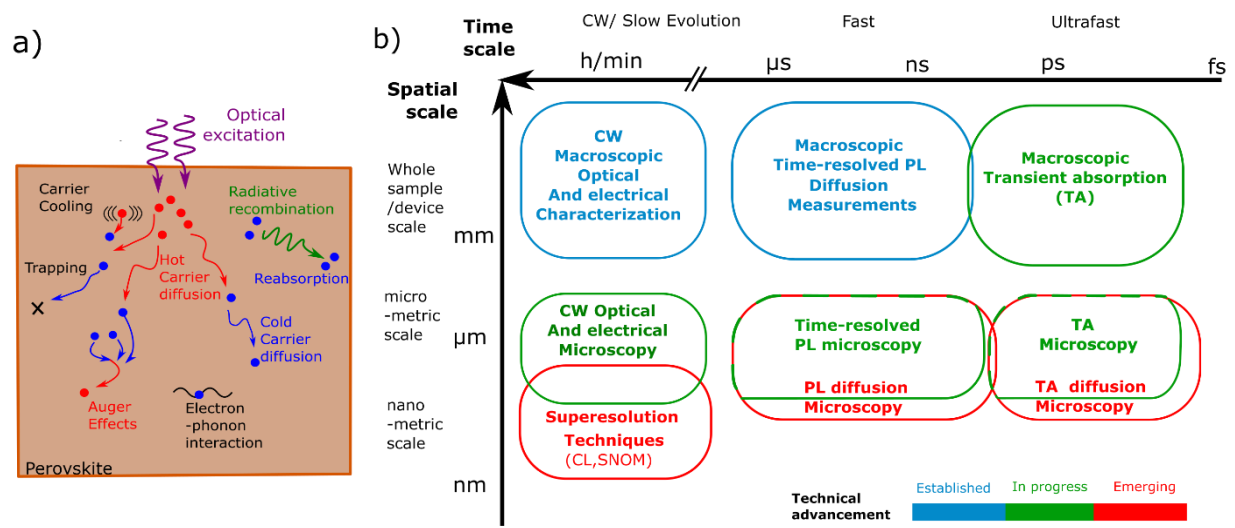

*Figure 1: Charge carrier processes in perovskites and the corresponding tools to study these effects. a) Schematic highlighting key photophysical phenomena related to carrier transport that can occur following photoexcitation of the perovskite. b) Table summarising the main experimental techniques mentioned in this Progress Report that are employed to study the recombination and diffusion of charges carriers in perovskites. These techniques are sorted as a function of three criteria: their level of technical advancement, and their relevant length and time scales accounting for typical resolution limits. CW denotes continuous wave illumination, CL is Cathodoluminescence and SNOM is Scanning Near-field Optical Microscopy.*

# 1. General recombination and transport considerations

When an optical beam impinges on a perovskite material, a cascade of physical effects will occur over different timescales (see summary schematic in Figure 1 a). If the energy of the photon exceeds the bandgap, this begins with the creation of hot carriers that can rapidly diffuse over hundreds of nanometres (within hundreds of femtoseconds[27]). These carriers will also interact with the lattice (carrier-phonon interaction), leading to a progressive cooling of the carriers down to the lattice temperature and the band edge normally within a few picoseconds[28,29]. Once at the band edge (cold), carriers continue to spatially diffuse through the perovskite material until they undergo a recombination event or are extracted from the system (e.g. at a device contact); these processes occur over the tens of picoseconds to microsecond time scales. Recombination can occur via several pathways, either radiatively through emission of a photon, or non-radiatively, such as through carrier-carrier interactions[30,31] or carrier trapping[15,31,32]. The emitted photon may then escape the film or be re-absorbed by the material, generating another photo-excited electron and hole at the band edge that can undergo further diffusion and/or recombination processes (i.e. photon recyling[33,34]). Each of these processes impacts the time- and spatially dependent population of charge carriers and can therefore in principle be probed through time-resolved (TR) spectroscopy techniques (summarised in Figure 1b).

To understand these competing processes, one must first consider the rate equation employed to express the local evolution of carrier density $n(r,t)$, which depends both on time $t$ and 3D spatial coordinate $r$. Most halide perovskites investigated to date are near-intrinsic semiconductors in which the number of electrons and holes are comparable, and thus $n$ represents the density of either electrons or holes. Upon photoexcitation, $n$ follows the rate equation [6,34]:

$$\frac{dn}{dt} = G - k_1 n - k_2 n^2 - k_3 n^3 + D\nabla^2 n \qquad (1),$$

where $G(r,t)$ represents the charge generation term (through optical or electrical injection), $k_1$ is the non-radiative recombination coefficient due to carrier trapping, $k_2$ represents the radiative recombination of electrons and holes and $k_3$ represents multiple-carrier (non-radiative) Auger processes that have an appreciable impact at high carrier densities (typically above ~$10^{18}$ cm$^{-3}$). The term $D\nabla^2 n$ corresponds to the ambipolar diffusion of charge carriers, quantified through the diffusion coefficient $D$. Note that equation (1) is an approximation that is valid for many 3D perovskites, such as MAPbI$_3$ (MA= CH$_3$NH$_3$), while a different system of equations[15] will be required in materials where other effects strongly impact the transport such as competing excitonic[31,35,36] effects or trapping/detrapping equilibria[37,38]. Optical spectroscopy measurements, such as TRPL or transient absorption (TA), are practical tools to evaluate the temporal and spatial evolution of the carrier density (see Figure 1b), $n$, and the different processes that will influence these populations (cf. Figure 1a). One can use for instance fluence-dependent measurements to isolate processes in different recombination regimes, by moving from a trap-limited regime (low carrier density) in which non-radiative trap recombination dominates, to a high carrier density regime in which traps are filled but Auger effects become increasingly important[15].

In many reports, the diffusion term in equation (1) is considered negligible or averaged out, though this assumption is not always valid. In particular, the in-depth diffusion[6,34] will inevitably play an important role in perovskite materials (regardless of the excitation beam size); the high absorption coefficient[39] means that carriers are predominantly generated in the first ~50-100 nm of the sample for most optical excitation wavelengths typically employed in time-resolved measurements, leading to subsequent diffusion of carriers further into the film (typically ~500 nm thick) or crystals (>>10 μm thick). Due to the high diffusion coefficients (~1 cm$^2$s$^{-1}$) measured in perovskite materials, charges will have distributed uniformly throughout a film after a few nanoseconds [40,41]. In that case, the diffusion term may indeed then be negligible in macroscopic measurements at longer time scales. In general, this term will need consideration, particularly over the early time scales and when considering local effects.

A variety of macroscopic optical measurements have been used to measure the carrier mobility and hence diffusion properties of carriers in halide perovskites. A straight-forward method to optically measure the diffusion coefficient and the diffusion length is to model the decrease of the TRPL lifetime in the presence of a quenching layer using equation (1) [6,8,42]. These diffusion parameters have also been extracted using time-resolved microwave photoconductance[35,43,44] and optical-pump terahertz-probe[30] spectroscopies [45]. Using these macroscopic techniques, charge-carrier diffusion lengths under solar-relevant carrier densities between several hundreds of nanometres[42,46,47] to several micrometres[8,48,49] have been reported, highlighting the broad distribution of values between different perovskite samples that depend on degrees of film quality, composition and passivation.

We emphasise here that diffusion properties are influenced by both competing physical processes and experimental factors. Thus, measured diffusion coefficients and lengths vary with excitation conditions or temperature. These values will also differ for each carrier type for cases where carrier transport is not ambipolar. In addition, diffusion is a microscopic process that depends strongly on the local quality and structural properties of the material, and thus such macroscopic measurements won't capture important local variations. Indeed, the values for diffusion coefficient determined macroscopically are spatially averaged and may differ significantly from values determined locally. For instance, local variations of the perovskite microscopic properties such as trap density and grain boundaries can lead to distinctly different local and macroscopic diffusion coefficients and lengths (further discussed below). Therefore, $D$ in Equation (1) needs to be considered a function of spatial coordinate, i.e. $D(\mathbf{r})$. Another source of discrepancy between the local and macroscopic diffusion coefficient values arises from contributions by photonic transport in which emitted photons propagate some distance and are reabsorbed in other regions of the sample (cf. Figure 1a)[34,50–52]. To account for photonic transport in equation (1), the charge generation term *g(r,t)* must include a contribution from photons that are locally re-absorbed, which will in turn depend on a number of parameters including the distance the photon has travelled (dictated by the emission wavelength and absorption coefficient) and the angular distribution of emission accounting for waveguiding and photon outcoupling effects. This photonic effect can modify the apparent diffusion coefficient with respect to the purely electronic value. In polycrystalline 3D perovskite films, studies have shown that the photonic transport is non-negligible, while the purely electronic diffusion coefficient may be orders of magnitude lower than the apparent one[34]. In 2D perovskite single crystals, Gan *et al*. have shown that this photonic effect can enable the out of plane transport of carriers between lead halide sheets[53], while purely excitonic diffusion in the plane is highly improbable due to the dielectric confinement effect of the organic layers.

## 2. Time-resolved PL microscopy to measure diffusion over micrometers and nanoseconds

TRPL microscopy is the most widely used and versatile tool to measure local diffusion properties. One advantage over other local spectroscopic techniques is that TRPL systems are compatible with measurements under realistic device-relevant excitation conditions such as with low photo-excited charge carrier densities relevant to solar cell devices (~$10^{15}$ cm$^{-3}$, fluences of < 0.1 μJ/cm$^2$/pulse in visible excitation wavelengths), and can easily cover many relevant temporal decades (10 ps to 100 μs). To quantify diffusion, a classical TRPL microscope needs to be adapted to spatially separate the excitation and collection areas (see Figure 2a). This can be achieved by using a local excitation and a widefield collection of the emitted light[54–56], or by raster scanning the collection (in a confocal configuration) while the local excitation is fixed[40,57]. Using either of these configurations, one can measure the evolution of the TRPL intensity as a function of distance from the excitation location. To study the diffusion process and extract the diffusion coefficient, the obtained TRPL results are fitted using an appropriate approach which also takes into account the finite diffraction-limited spatial excitation profile. Several fitting or simulation methods have been proposed, all based on variations of equation (1)[41,58,59]. This process generally requires a preliminary estimation of injected carrier density $n$ and of the value of the $k_1$ and $k_2$ coefficients[41]. Indeed, each type of recombination will also decrease the local $n(r,t)$ and therefore impact the measurement of diffusion, for example by generating an apparent broadening of the PL profile as a function of time that must be distinguished from true carrier diffusion. An example of a TRPL approach is the work from Tian *et al.* [7] (seen in Figure 2b-d), where an intense PL signal is observed several microns (point A) away from the excitation location in perovskites single crystals confirming the long (i.e. micrometre scale) diffusion length in such materials. They then quantified the TRPL intensity originating from diffusion (in red in Figure 2d) by removing the contribution from waveguided (WG) photons and applying a fitting process. Using an excitation fluence of 5.3 µJ.cm$^{-2}$ at 1 MHz, they evaluated the diffusion coefficients to be ~ 1 cm$^2$s$^{-1}$ for MAPbBr$_3$ and ~2 cm$^2$s$^{-1}$ for MAPbI$_3$ crystals and the corresponding diffusion length under these excitation conditions to be ~1 µm. In another study, the characteristic Gaussian width $\sigma(t)$ of the measured carrier distribution was shown to follow a classical diffusion law[40]:

$$\sigma^2(t) = \sigma^2(0) + 2Dt \qquad (2)$$

where $t$ is the time after the optical excitation of a local area of the sample, and $\sigma(0)$ is the initial Gaussian spatial distribution. Stavrakas *et al.* used a two-photon photoluminescence microscopy configuration[19,60] to study the diffusion process in the bulk of a MAPbBr$_3$ crystal. While the mean value of the diffusion coefficient (~1.4 cm$^2$s$^{-1}$, yellow dash line in Figure 2e) is similar to the value cited above, this study reveals that there is a large spread of local diffusion coefficients in both lateral dimension and depth, which is attributed to local heterogeneities in trap densities and defects in the crystal[40] (see Figure 2e).

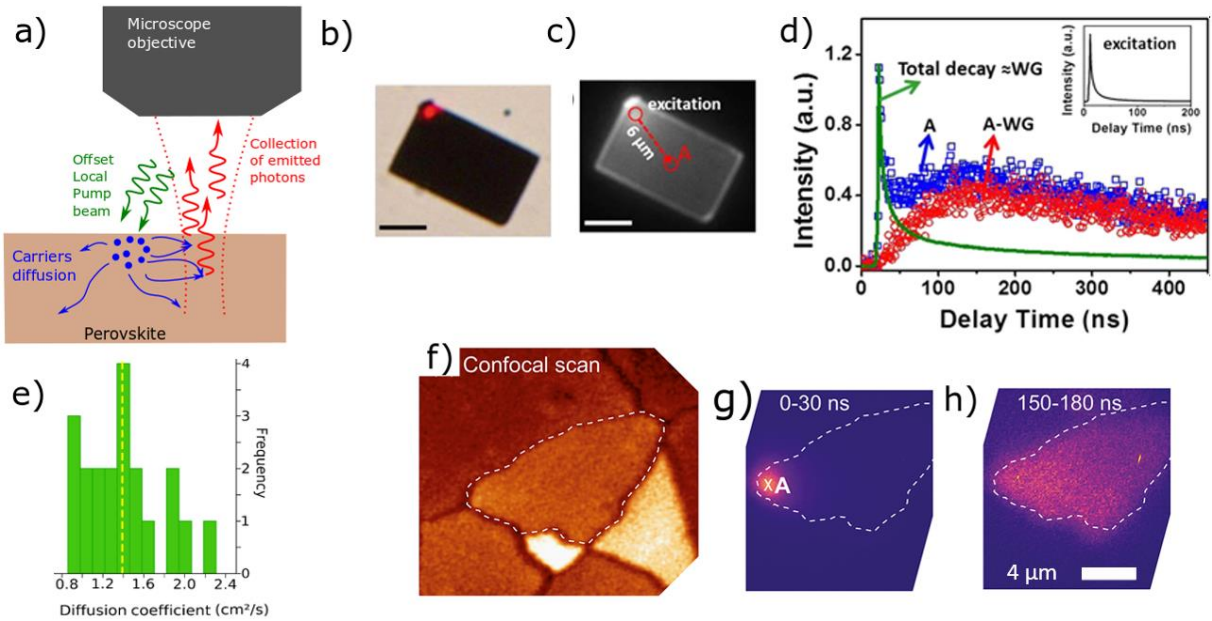

*Figure 2: TRPL microscopy to assess the local diffusion of carriers. a) Schematic detailing the use of TRPL microscopy to probe the diffusion of carriers. b-d) Carrier diffusion over several microns within a single crystal probed by TRPL microscopy. The TRPL signal is analysed (d) and modelled against equation (1) to decouple the two major contributions: photon waveguiding (WG, in green) and carrier diffusion (in red) for a distance of 6 microns between the excitation and the point of collection of the PL signal (see c). Adapted with permission from [7]. Copyright (2019) American Chemical Society. e) Distribution of local diffusion coefficients measured within the bulk of MAPbBr$_3$ single crystals using a two-photon TRPL microscope [40]. f-h) TRPL microscopy measurements exhibiting the diffusion of carriers within a MAPbI$_3$ large grain sample, showing that the lateral spreading of the PL profiles due to carrier diffusion as a function of time is confined to one grain, while the grain boundaries (visible in f or with the dashed lines in g and h) inhibit the diffusion of carriers. Adapted with permission from [41]. Copyright (2019) American Chemical Society.*

Although carriers follow a classical diffusive behaviour in perovskite single crystals, this does not appear to be the case in polycrystalline films in which grain boundaries are prevalent. Ciesielski *et al*[41] used TRPL microscopy to show that grain boundaries in the MAPbI$_3$ films they considered act as barriers inhibiting carriers from diffusing efficiently over large distances in polycrystalline perovskites films (see Figure 2 f-h). However, carriers could efficiently diffuse within grains (within dashed lines in Figure 2 g-h) as a function of time, and the diffusion coefficient they obtained (~1 $cm^2s^{-1}$) was similar to those measured in other single crystals studies. Using similar methods, other groups have reported diffusion coefficient values that are two orders of magnitude lower in polycrystalline perovskites thin films than single crystals, despite long carrier lifetimes in the former[6,34]. This is likely because these microscopic measurements probe several grains together, hindering any observable broadening with time of the PL spatial profile. This effect can be particularly exaggerated if the lateral size of grains is smaller than the optical beam diameter (~ 600 nm), as it is for state-of-the-art mixed halide triple cation perovskites samples, for example[13]. Sridharan *et al.* [58] have shown that the diffusion coefficient in perovskite polycrystalline films (with various compositions) decreases significantly over time, with values similar to single crystals on the picosecond timescale (~1 $cm^2s^{-1}$) but decreases to ~$10^{-3}$ $cm^2s^{-1}$ after hundreds of nanoseconds. This effect could be explained by the confinement of carriers within grains allowing fast initial intra-grain transport but the inhibition of the larger-scale diffusion of carriers at longer decay times. Alternatively (or additionally), carrier trapping [15,61] of one or both carrier species will detrimentally influence carrier transport, an effect which may be more pronounced at longer times (and/or at further distance from the excitation spot) when carrier densities are lower and thus traps may not be locally saturated by carriers. In any case, these studies highlight that the relationship between the carrier lifetime and the diffusion process is complex in polycrystalline materials; in particular, one cannot assume that a longer PL lifetime always implies a more efficient diffusion of charges across the perovskite thin film.

## 3. Studying inter- and intragrain energetic transport with pump-probe microscopy

TRPL microscopy is a practical tool for evaluating diffusion properties but it suffers from several limitations. Such measurements exclusively probe the radiative recombination of electrons and holes, meaning that only ambipolar diffusion can be measured. As a consequence, TRPL is not able to directly probe non-radiative recombination events that have an important impact on diffusion[58]. When electrons (holes) are trapped locally, it is difficult to study the diffusion of remaining free holes (electrons) using TRPL because such transport will not necessarily result in photon emission. Due to the high absorption coefficients of perovskite samples, PL is also mostly sensitive to radiative events occurring near the surface because photons emitted deeper in the crystal have a higher probability of being reabsorbed before escaping and ultimately not leading to a photon emitted externally from the film (cf. Figure 1a).

Transient absorption (TA) spectroscopy is one of the fundamental pump-probe techniques that has become ubiquitous in the photophysical characterisation of semiconductors, particularly organic[62] and, more recently, halide perovskite systems[33,63–65]. Transient absorption microscopy (TAM) is a powerful variant used to spatially image the transient energetic distribution of excited species within a sample, with time resolution down to the femtosecond regime, and with spatial precision that can far surpass the diffraction limit[66–68]. This information comes without relying on the radiative efficiency of the sample. Of the possible modalities of TAM, the two most widely adopted for diffusion studies are fixed pump/scanned probe (as illustrated in Figure 3a) and fixed pump/wide-field probe. In both cases, pump and probe bandwidths can vary but spectral selectivity will become limited by the time-bandwidth product when pulses are sufficiently temporally compressed[69]. As a transmission measurement, TAM yields a convolution of the excited state dynamics throughout the whole sample thickness and is therefore complementary to more-surface-sensitive PL measurements. Combining both measurements in situ, Simpson *et al*. detect significant variation in PL maps despite TAM revealing a relatively uniform carrier population, emphasising the need to locally isolate variations in recombination kinetics and transport in light of relevant experimental factors[70].

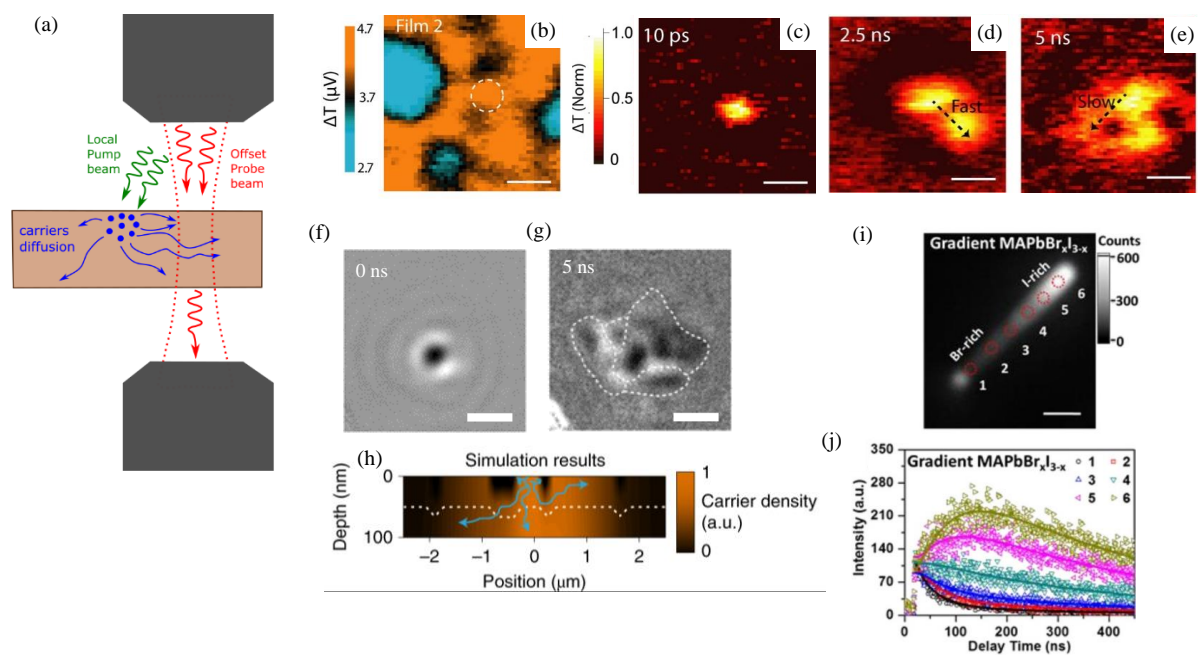

*Figure 3: Energetic transport visualised with pump-probe and PL microscopy.* *(a) Schematic of Transient Absorption Microscopy (TAM) showing the local photogeneration of excited species. The transport of excited species can be tracked by scanning a local probe (shown) or using a wide-field probe. (b) Wide-field TAM image (630 nm pump, 755 nm probe) at pump-probe delay time of 10 ps, revealing MAPbI$_3$ sample morphology. The white dashed circle indicates the location of the pump spot. (c)-(e) Diffusion across grain boundaries, visualised with TAM with local pump (white circle in (b)) and raster-scanned probe at delay times of 10 ps, 2.5 ns and 5 ns, tracking carrier diffusion between grains. Panels b-e are adapted from* [71]. *Copyright (2018) American Chemical Society. (f) & (g) Stroboscopic scattering microscopy images at 0 and 5 ns pump-probe delay (scale bars 1 µm). Maps of magnitude and phase of scattering contrast yield information about lateral and depth-dependent carrier density. Negative contrast (black) represents a photoexcited population within ~30 nm of the film surface and positive contrast (white) delineates considerable carrier density at a depth of ~50*

*nm. (h) 2D simulation of diffusive carrier population accounting for influence of observed morphological boundaries. Panels f-h are adapted from [72]. Copyright (2019) Springer Nature. (i) Map of TRPL intensity from a MAPbBr$_x$I$_{3-x}$ nanowire (scale bar 2 μm). (j) TRPL decays from local spots on the nanowire, labelled in (i), showing a ~150 ns rise time at the iodine-rich region, caused by the energetic funnelling of carriers along the bandgap gradient. Panels i & j adapted from [73]. Copyright (2017) American Chemical Society.*

Katayama *et al.* used spatially resolved TA measurements to identify local variations in the excited carrier relaxation in mesoporous MAPbI$_3$ films[74]. Early work on these systems by Guo *et al.*[75] utilised TAM to observe the diffusive expansion of a carrier population with ~50 nm spatial precision. In calculating exceptionally high diffusion coefficients for solution-processed films at the time (up to 0.08 cm$^2$s$^{-1}$, 1 ns after excitation), they identified grain boundaries as the discrepancies between their own measurements of intragrain diffusion, and literature values obtained via quenching contact methods[8]. Recent TAM developments have subsequently enabled direct examination of diffusion across grain boundaries. Snaider *et al.* measured the diffusion coefficient for transport across a boundary to be ~0.12 cm$^2$s$^{-1}$, compared with ~0.20 cm$^2$s$^{-1}$ in the grain interior[71]. On identifying the grain morphology (Figure 3b) they additionally observe disparate rates of carrier transport between neighbouring grains (see Figures 3c-e). The moderate suppression of transport across boundaries is consistent with TRPL studies[76] and again reiterates the limitations of bulk measurements that will spatially average out such behaviour. The observation of spatially varying diffusion motivates local sampling of diffusion coefficients to build up statistics of transport properties[77,78]. However, these results also underline the comparatively benign effect that grain boundaries have on lateral carrier diffusion in perovskites, in contrast with polycrystalline silicon whose inter-grain mobility can be crippled by deep traps formed at boundaries[79–81].

Delor *et al.* resolve both lateral and vertical motion of carriers in polycrystalline MAPbI$_3$ using a state-of-the-art scattering-based pump-probe microscope.[72] Figures 3f and g respectively show the scattering signal portraying the initial pump-induced carrier distribution and the subsequent dispersal after 5 ns. The depth-dependent phase of the interferometric signal allows the authors to identify and model (Figure 3h) a carrier population which diffuses away from the sample surface when navigating grain boundaries. Despite pump and probe 1/*e* penetration depths of only ~50-70 nm in MAPbI$_3$, the technique provides unique insight into the vertical distribution of carriers that would not be available in conventional local PL or TAM measurements. Furthermore, another dimension has been added to the complexity of grain boundary effects: lateral transport also depends on depth, and the potential barriers presented by boundaries vary both in width and depth; this is also seen through two-photon measurements in MAPbBr$_3$ single crystals, where transport varies significantly at different depths in the crystal[40]. It remains to be seen whether such vertical heterogeneity in transport will be present in state-of-the-art perovskite solar cell absorber films.

Optical spectroscopy can be complemented by other techniques to study local photophysics under the influence of compositional heterogeneity, such as the spatially varying distribution of halide ions in alloyed perovskite materials[82,83]. Tian *et al.* combine energy-dispersive X-ray spectroscopy (EDS) with time-resolved PL microscopy to visualise the funnelling of carriers along an energy gradient in single-crystalline perovskite nanowires[73]. The energy funnel is set up by a varying ratio of halide ions along the MAPbBr$_x$I$_{3-x}$ wire, such that excited carriers will cascade down to the lower bandgap, iodine-rich region. Increased PL emission is observed from the iodine-rich end of the nanowire, fed by carriers transferring from the bromine-rich end on on which the local TRPL signal is rapidly quenched (Figure 3i). A comparison between the TRPL rise time (up to 150 ns, as shown in figure 3j) and bandgap variation along the wire allows the authors to decouple carrier motion due to energetic funneling from carrier diffusion. The calculated diffusion coefficient of ~1.2 cm$^2$s$^{-1}$ lies between values measured for pure MAPbBr$_3$ and MAPbI$_3$ single crystals[7]. Carrier transfer to the surface of perovskite thin films can be similarly driven by light-induced halide segregation[84,85] and such energy funnelling is common in 2D perovskites in which phases of different bandgap coexist[86]. This funnelling is exploited in LEDs based on quasi-2D layered perovskite[23], and on perovskite nanograins[87], which are respectively engineered to concentrate and confine injected carriers, yielding high radiative efficiencies. In inorganic 2D materials, strain-induced bandgap engineering[88] has been employed to induce charge funnelling, and similar behaviour has been explored in halide perovskites[89].

## 4 Ultrafast Processes & Hot Carrier Diffusion

Recent applications of TAM are uncovering the increasingly complex role that ultrafast processes play in dictating local carrier recombination and transport. The ability to spatially resolve carrier cooling processes is one such valuable insight, not least because hot carrier solar cells have the potential to in principle push the theoretical power conversion efficiency limit to 66%[90,91]. Hot carriers in MAPbI$_3$ have been shown to exhibit remarkable long-range (~600 nm) diffusion on picosecond timescales[27]. The hot carrier population undergoes rapid expansion, with the apparent diffusion coefficient exceeding 450 cm$^2$s$^{-1}$ upon photoexcitation at approximately twice the bandgap energy (see Figure 4a). The diffusion coefficient decays on the tens of picoseconds timescale to 0.7 cm$^2$s$^{-1}$, a value in line with reported diffusion coefficients for cold carriers[7]. This timescale is attributed to the slow formation of polarons which are predicted to affect charge transport[92,93]. Studying MAPbI$_3$ films formed via different fabrication methods, Sung et al. [66] utilised TAM with higher temporal resolution to observe the ballistic transport of carriers on an even shorter timescale. The super-diffusive motion of the carrier population occurs over distances of up to 150 nm within 20 fs of photoexcitation (see Figure 4b), and has a spatial variance, $\sigma^2$, that grows quadratically with time unlike the linear dependence of a standard diffusive process represented by equation 2. Together, these works emphasise that ballistic and quasi-ballistic transport occur over distances that are significant relative to device dimensions (~500 nm thickness) and could directly contribute to carrier extraction at contacts. Moreover, while carrier-carrier interactions are seen to mitigate ballistic diffusion at high carrier densities, the transport-limiting factor at typical solar fluence is most likely carrier-lattice interactions dictated by energetic disorder within the material[66,77]. While grain boundaries in polycrystalline films undoubtedly contribute to this disorder, it is as yet unclear whether the increased kinetic energy of ballistic and hot carriers allows them to more easily negotiate the potential barriers presented by boundaries.

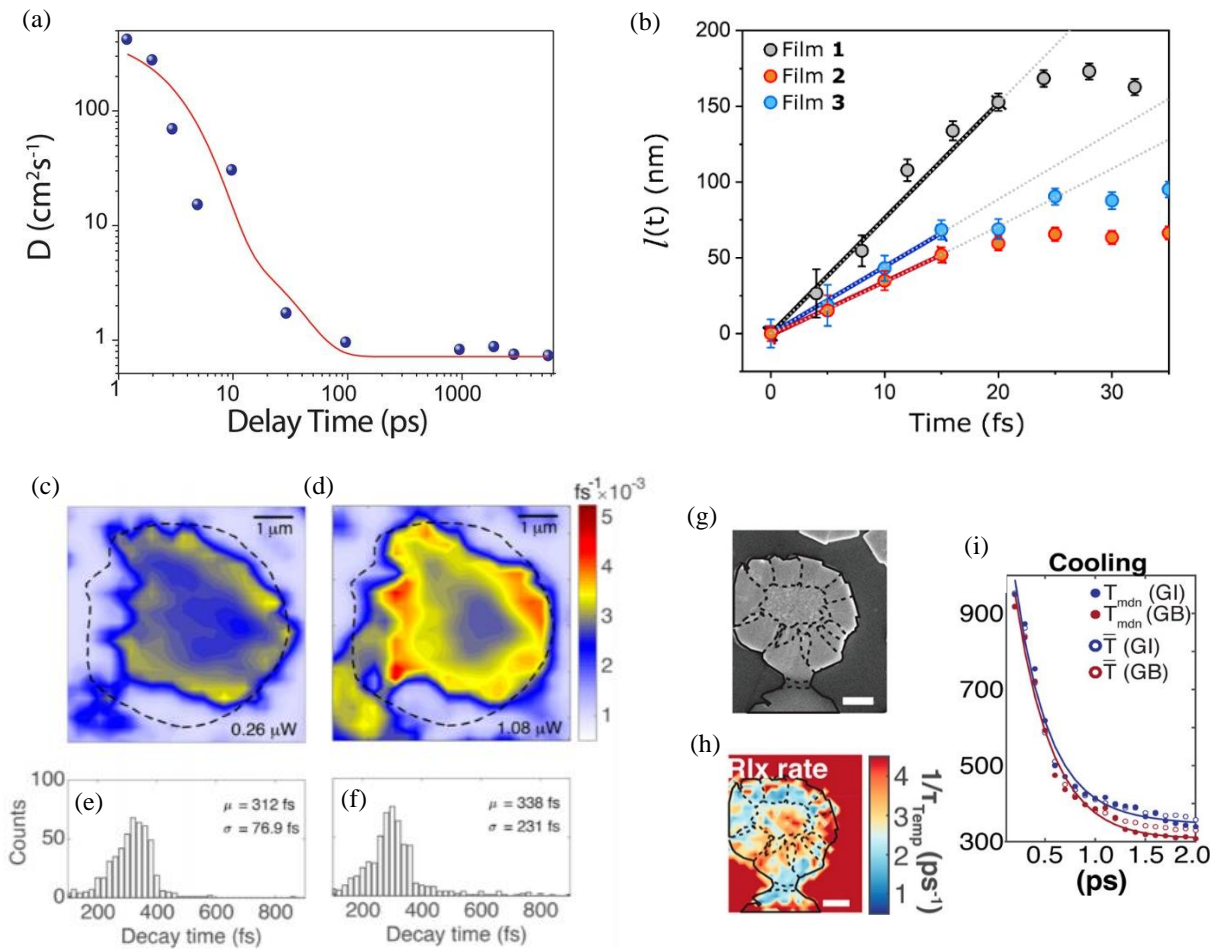

*Figure 4: Femtosecond TAM measurements to study hot carrier dynamics and ballistic motion. (a) Evolution of diffusion coefficient during slow carrier cooling process in polycrystalline MAPbI$_3$. Adapted from [27]. Copyright (2017) American Association for the Advancement of Science. (b) Variation of ballistic carrier transport length, l(t), during ultrafast*

*expansion in MAPbI₃ films with different morphology. Lateral diffusion of 150 nm within 20 fs is observed in the film showing least energetic disorder (black line), while ballistic expansion is curtailed in more heterogenous films (blue and red lines). Adapted from [66]. Copyright (2019) Springer Nature. (c) & (d) Spatially resolved carrier cooling rate maps of polycrystalline MAPbI₃ samples at excitation powers of 0.26 µW and 1.08 µW. (e) & (f) Histograms of pixel decay times at corresponding excitation powers. Increased fluence leads to larger average cooling rate and an increased spread with more locations exhibiting extremely slow cooling. Panels c-f adapted from [91]. Copyright (2019) American Chemical Society. (g) SEM image of MAPbI₃ crystallite with grain boundaries highlighted. (h) Map of calculated hot carrier relaxation rate with outlines of grain structure overlaid (scale bars 1 µm). (i) Evolution of carrier temperature at grain interiors (GI) and grain boundaries (GB), showing accelerated cooling at boundaries. Panels g-i adapted from [94]. Copyright (2019) American Chemical Society.*

In 3D perovskites such as MAPbI$_3$ it is widely reported that hot carriers undergo slow cooling at high excitation densities due to processes such as the hot phonon bottleneck effect[29,63]. Using TAM, Nah *et al.* reveal a broad distribution of spatially varying carrier cooling rates across micron-scale MAPbI$_3$ crystals upon above-bandgap excitation (see Figures 4c & d)[91]. They observe that the mean cooling rate decreases with increasing excitation fluence (see Figures 4e & f), matching earlier macroscopic studies, but additionally find that the variance in local cooling rates significantly increases with fluence, indicating that the onset of the hot phonon bottleneck effect is changeable spatially across the sample. Larger crystal domains are observed to facilitate faster cooling, suggesting that carriers that experience localisation or confinement exhibit a stronger hot phonon bottleneck effect[95]. Interestingly, Jiang *et al.* image polycrystalline perovskite particles with SEM-correlated TAM (see Figures 4g & h) and report faster cooling rates at grain boundaries compared with grain interiors, which they propose is due to an energy exchange process mediated by sub-bandgap states (see Figure 4i)[94].The proposed trapping and de-trapping of photoexcited holes will likely saturate at moderate excitation densities[15], such that small grains still experience a prolonged elevated carrier temperature due to confinement, despite an increased surface-to-volume ratio. Additionally, such grain boundary interactions (effectively scattering processes) will likely obstruct carrier diffusion across grain boundaries, providing another explanation for the differing intra- and inter-grain diffusion coefficients reported earlier. It is apparent that initial carrier temperatures and the time evolution of relaxation processes can be directly linked to the early time diffusion of excited species and their eventual fate. Such measurements therefore provide an alternative lever with which to study transport and uncover valuable information about the local energetic landscape of perovskite materials.

In the context of optical studies, it should be noted that TAM could be particularly susceptible to experimental artefacts and ambiguities, and thus analysis must be performed with extreme care. For instance, acquiring accurate steady state absorption spectra of perovskite films is challenging due to light scattering, and heterogeneity in sample morphology and optical density[59]. This issue extends to TA measurements where the transient transmission signal is a convolution of time-dependent changes in absorption and reflectance[96]. The relatively high refractive index of perovskites poses additional problems in that transient reflectance contributions can dominate the observed TA signal[63,97]. To obtain a rigorously representative picture of carrier diffusion, one needs to locally identify the pump-induced change in the full complex refractive index, covering both variations in absorption and reflectance[98]; achieving this on the micro- or nano-scale will be challenging but may prove to be very fruitful.

Artefacts in transport behaviour caused by spot-to-spot morphological variation can be accounted for by measuring at multiple locations[66] but this procedure cannot mitigate signals which arise due to optical effects linked to the pump and probe beams. Coherent artefacts can afflict TA signals on very short timescales and are the result of interference interactions between the ultrafast pump and probe pulses[99,100]. Since TAM studies of early-time ballistic transport are already being carried out on femtosecond timescales, it is important to be able to identify and correct for these effects[69]. A focussed pump beam, as is commonly used for TAM diffusion measurements, can create a highly localised refractive index variation which acts as an aperture for the pursuing probe, complicating the issue. The focussed excitation can also cause non-linear optical effects such as two-photon absorption which influence pump-probe measurements carried out at high fluence[101]. There is therefore additional motivation to move TAM studies of diffusion to lower fluence, on top of the already appreciable risk of sample deterioration that is reported under intense pump beams[91].

## 5  Outlook

### i.  *Gaining further insight into the nature of local heterogeneities*

From the atomic-scale up to grain-sized domains, our understanding of what drives spatial heterogeneity in carrier dynamics is limited. While optical microscopy techniques are suitable to probe the presence of such heterogeneities, for example local surface morphological features or indirect information on local trap states, they cannot provide information about the physico-chemical nature or mechanism that induces these properties. To address this issue a multimodal approach could be adopted, where local optical probes are correlated with nanoscopic probes of structural and chemical information. These local characteristics of the perovskite (e.g. stoichiometry, doping, crystalline phase, defect density) will directly impact local recombination and transport. The structural and compositional nature of the grain boundaries and the prevalence of multigrain crystalline order also remain uncharacterised and their effects present additional questions. For such studies, suitable structural microscopy techniques[10] that could be applied in conjunction with optical methods include nanoscale X-Ray diffraction[102] and/or electron microscopy diffraction techniques.

Cathodoluminescence[103] (CL) has been effactually used to study the luminescence properties of materials such as GaAs[104] and GaN[105,106] on the nanoscopic scale. Since CL detectors are often incorporated into scanning electron microscope (SEM) systems, *in situ* correlation of luminescence with surface morphology from SEM is straight forward[107,108]. The application of CL to inorganic perovskites has yielded luminescence maps with ~10 nm spatial resolution[109] (see Figure 5 a-b). However, the high electron beam doses required (typically in vast excess of the damage thresholds for MAPbI$_3$ of 100 electrons/Å$^{2[110,111]}$) are irreversibly destructive to hybrid organic perovskite materials which are in general very beam sensitive. Transient luminescence can also be probed on picosecond timescales (comparable to fast TRPL measurements acquired with streak cameras) using time-resolved CL[50,104]. Complementary TRPL and TRCL measurements have been employed to investigate the heterogeneous formation of multidimensional Ruddlesden-Popper perovskites[109]. By varying the voltage in a CL measurements, one can also controllably excite the perovskites at different depths to monitor the effects of reabsorption on the PL signal[50]. If CL and TRCL collection can be continually optimised for lower electron doses (available via technical advancements in pulsed excitation, for example) then this technique promises unrivalled spatially resolved luminescence as part of a powerful multi-modal toolbox.

Photoemission electron microscopy (PEEM) measurements can identify local sub-bandgap states on the nanoscale, thus allowing one to directly image non-radiative recombination centres in the sample. In addition to identifying trap states, PEEM is also suitable for imaging carrier motion in semiconductor materials[112,113] at the very local scale, since its spatial resolution is only limited by electron optics (~10 nm). Man *et al.* demonstrate the spatial and temporal resolution of time-resolved PEEM in their study of charge transport across a staggered-gap 2D heterostructure formed by an InSe/GaAs interface, allowing visualisation of charge flow across the heterojunctions[103]. Our group has recently applied PEEM to polycrystalline perovskite films to identify local surface trap clusters[114]. Figure 5c shows a PEEM image of local hotspots of photoemission from within the mid-gap of the perovskite band structure, which represent trap-rich local sites (white spots). These images reveal clusters of defects typically less than 100-nm in size that are in general smaller than morphological grains (of order hundreds of nanometres) that are resolved in a PEEM map using a higher energy probe (Figure 5d). Such nanoscale measurements of the sub-gap states give complementary information to the carrier transport and recombination measurements discussed above, and rich information will come from multimodal correlations between these measurements and optoelectronic, structural, morphological and chemical properties. Further, time-resolved PEEM measurements will allow one to assess the impact of these states on carrier diffusion on the nanoscale, complementing local TRPL and TAM measurements.

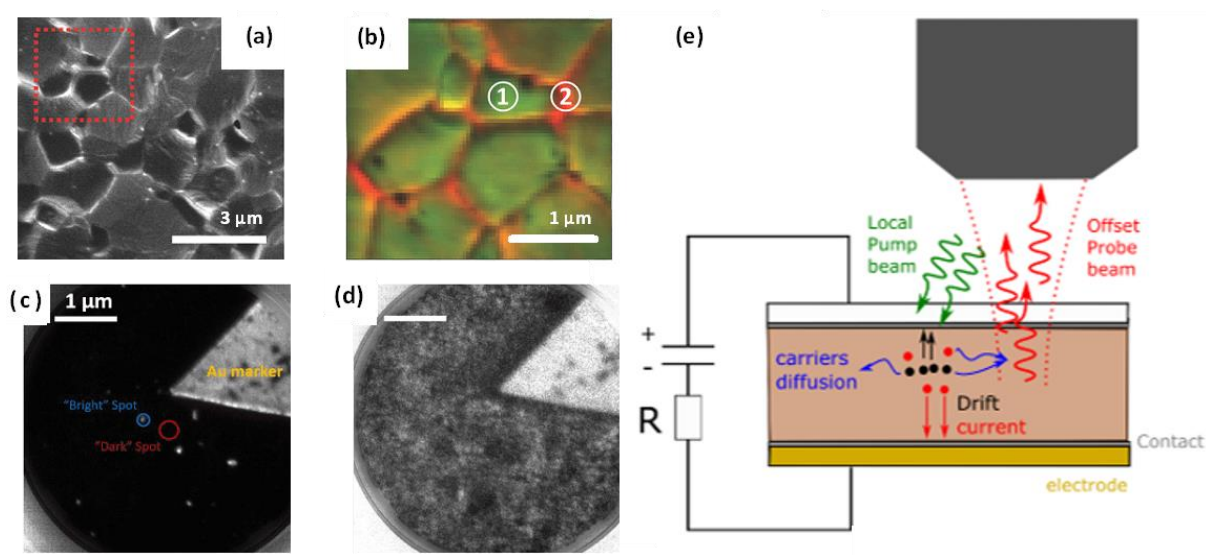

*Figure 5: Cutting edge experimental developments that could further enhance our understanding of carrier recombination and transport in halide perovskites. (a) SEM image of an inorganic polycrystalline perovskite film and (b) complementary CL map of the region highlighted with a red dashed line in (a). Each colour map in (b) corresponds to a different wavelength range: 530–590 nm (green) and 590–640 nm (red). Adapted from* [109]. *Copyright (2017) Wiley. Photoemission electron microscopy (PEEM) images of a polycrystalline perovskite film ($Cs_{0.05}MA_{0.17}FA_{0.78}PbBr_{0.17}I_{0.83}$) probed with (c) 4.65 eV pulses and (d) 6.2 eV*

*pulses. The latter high energy probe predominantly photoemits valence band electrons, revealing the surface morphology of the film. The low energy probe photoemits exclusively from mid-gap states, revealing discrete defect-rich locations on the sample surface. The Au particle in the image is a fiducial surface marker. Adapted from*[114]*. Copyright (2018) Institute of Electrical and Electronics Engineers. (e) Schematic of optical microscopy in operando device measurements probing diffusion under applied bias.*

### ii. *Studying carrier photophysics in operating devices*

It is evident that microscopic studies of carrier dynamics must be extended to devices (summarised in schematic in Figure 5e), in which excited species can experience markedly different conditions from those typically encountered during idealised spectroscopic measurements. For instance, the addition of a device contact creates an interface with the perovskite layer which has consequences for local transport properties that are not yet understood. As mentioned earlier, early time diffusion will homogenise the charge distribution within each grain (size ~ 500 nm) within a few nanoseconds. Beyond this, the presence of extraction layers will progressively modify the charge distribution, either because of additional non-radiative surface recombination events[6,115–117] or because of the depletion of a particular type of charge at each end of the device[118]. It is also important to understand why extraction efficiency is seen to vary spatially in some devices[119], despite relatively homogenous intragrain diffusion[120], and surface-sensitive imaging tools will be particularly valuable for such investigations. Macroscopic transient reflectance measurements have been utilised to decouple bulk and surface recombination in both perovskite single crystals and thin films[121,122] Steady-state reflectance microscopy is a staple of morphological characterisation and has been utilised as part of multimodal apparatus[76] but, to our knowledge, differential reflectance microscopy has yet to be applied to perovskites despite being demonstrated on other materials[123]. Again, multimodal measurements between local electrical and optical measurements will provide complementary information to assess the interplay between local extraction and recombination, properties which have been shown to vary on the nanoscale[124,125].

Future studies should also focus on isolating the diffusion properties of electrons from those of holes[6]. To do so, one could exploit the local depletion of electrons (holes) induced by a selective extraction layer in order to measure the diffusion of holes (electrons). Tainter *et al.*[118] utilised this approach using a back contact architecture where electron and hole selective layers are positioned next to each other (rather than in a vertical stack), revealing particularly long hole diffusion lengths of 13 μm following removal of electrons. Such studies are crucial to better our understanding of individual carrier mobilities, but also in the design of advanced devices that could exploit such long and/or selective carrier diffusion behaviour.

Critical insight into how device operating conditions modify carrier dynamics could be gained from performing optical microscopy measurements under different bias conditions (e.g. short circuit, open circuit, maximum power point). When a voltage is applied to the device, a vertical drift of charge may occur due to any net electric field across the device; in this case, Equation (1) would need to be modified to account for drift motion of charges. One could use time-resolved optical microscopy to unveil how this drift competes or coexists with diffusion processes. By way of example, Yang *et al.* studied the dependence of carrier transport on the electric field applied across $MAPbBr_3$ microplates, observing a clear shift from diffusion-like to drift-like transport at higher fields[126]. The impact of ionic transport in the perovskite structure on the screening of the electric field felt by the charges must also be considered in such device-like measurements[127,128]. If this screening effect is strong, it could supress the drift current in perovskite solar cells, in which case diffusion would remain the dominant mechanism for the efficient transport of charges in devices even at high bias; this has important ramifications for optoelectronic device design. In LED devices, the voltages applied are much stronger (up to several volts) and, unless the device design is controlled to mitigate this, drift is likely the main charge transport mechanism. Such bias-dependent local measurements provide ways to distinguish these competing mechanisms and will be critical for further spectroscopic insights into real, operating devices.

Finally, the optical techniques described in the previous sections reveal rich information about carrier dynamics in perovskites but most are performed under excitation conditions that do not represent those found in operational device stacks. Efforts are needed to further develop the techniques to allow study of samples with excitation densities and carrier injection levels that best match realistic device operating conditions (for example, typical carrier densities of $10^{13}$-$10^{15}$ cm$^{-3}$[15] for solar cells and $10^{17}$-$10^{19}$ cm$^{-3}$ for LEDs). Numerous works on photovoltaic materials (particularly those employing pump-probe techniques) use excitation densities that are well above solar fluences, typically due to requirements for sufficient signal-noise ratios; such high carrier densities may also bring with it sample heating and degradation issues[91]. This is not to undermine the value of such high fluence and carrier-density-dependent measurements, which when performed carefully reveal valuable information that fit into our collective understanding. We reiterate that extracted transport and recombination properties need to be reported together with the excitation and bias conditions, and there should be active push towards including realistic device-like excitation conditions. As has become standardised for reports on device characteristics, the publication of carrier transport parameters under common excitation conditions would greatly simplify the comparison of perovskite materials and devices within the literature and would expedite our understanding of their limitations and, ultimately, their optimisation for end device use. The toolsets described above could provide a facile means to do so and, with continued ongoing developments, will continue to push the traditional experimental limits to extract critical information about these intriguing semiconductors.

**Conclusion**

We have explored the use of local time-resolved optical spectroscopy as a toolset to visualise carrier recombination and transport in halide perovskite materials. The versatility of these all-optical techniques reveals a plethora of competing photophysical processes including transport and recombination across multiple length and timescales. However, these processes need to be decoupled from each other as well as from experimental artefacts in order to isolate electronic transport processes, and the use of complementary techniques such as time-resolved photoluminescence and pump-probe microscopy methods can help to do so. The temporal evolution of the measured transport mechanisms has been discussed, with ballistic transport occurring at early times (picoseconds) followed by the progressive decrease of the diffusion coefficient at later times during carrier cooling. Local sample heterogeneities including traps and grains boundaries have an important inhibiting effect on carrier transport, and further understanding of the nature of these sites will be critical for our understanding and ultimately control of these heterogeneities. This motivates the use of multimodal techniques to assess how the local transport (optical and electronic measurements) relates to local chemical and structural properties, as well as further developing new measurement techniques that continue to push the limit on spatial and temporal resolution and/or provide complementary insight into the processes such as low-dose time-resolved cathodoluminescence and photo-emission microscopy measurements. Finally, measurements must move towards in operando device characterisation to truly understand how the recombination and transport properties are modified in operational devices under device-relevant excitation and bias conditions, which will in turn guide material and device design. The cutting edge technical developments made in this field will be applicable to a wide range of established and emerging semiconductor systems including during the development of new perovskite-inspired materials that may exhibit similar defect tolerance to the metal halide perovskites.

Acknowledgements

S. D. S. acknowledges support from the Royal Society and Tata Group (UF150033). G.D. acknowledges the Royal Society for funding through a Newton International Fellowship. S.M. acknowledges funding from an EPSRC studentship.

Conflict of Interest

SDS is a co-founder of Swift Solar Inc.